\documentclass[review=false]{jfp-epi}

\usepackage{jfp-cup2epi}

\usepackage{cleveref}
\usepackage{tikz}
\usepackage{listings}
\usepackage{color}
\usetikzlibrary{arrows, arrows.meta, automata, calc, shapes, positioning, fit}

\definecolor{darkblue}{rgb}{0,0,0.5}
\definecolor{darkgreen}{rgb}{0,0.3,0}
\definecolor{darkpink}{rgb}{0.4,0,0.3}
\definecolor{graygreen}{rgb}{0.3,0.5,0.3}
\definecolor{grayblue}{rgb}{0.2,0.2,0.6}
\definecolor{grayred}{rgb}{0.5,0.2,0.2}

\newcommand{\todo}[1]{}
\renewcommand{\todo}[1]{{\color{red} TODO: {#1}}}

\AtBeginDocument{%
  }

\journaltitle{JFP}
\cpr{Cambridge University Press}
\doival{10.1017/xxxxx}

\lefttitle{Trustworthy Runtime Verification via Bisimulation (Extended Experience Report)}
\righttitle{Journal of Functional Programming}

\totalpg{\pageref{lastpage01}}
\jnlDoiYr{2025}

\begin{document}

\newcommand{\copilot}{Copilot}
\newcommand{\velus}{V\'elus}
\newcommand{\copilotC}{\textsc{CopilotC99}}
\newcommand{\copilotTheorem}{\textsc{CopilotTheorem}}
\newcommand{\verifier}{\textsc{CopilotVerifier}}
\newcommand{\crucible}{\textsc{Crucible}}
\newcommand{\crucibleLLVM}{\textsc{CrucibleLLVM}}
\newcommand{\whatFour}{\textsc{What4}}
\newcommand{\crux}{\textsc{Crux}}
\newcommand{\listSet}[1]{\mathcal{L}(#1)}

\tikzstyle{nodeStyle}=[
    draw = black,
    thick,
    minimum size = 4mm,
    rectangle,
    rounded corners,
    inner sep = 2mm,
    align=center
]

\tikzstyle{edgeStyle}=[
    ->,
    align=center
]

\title{Trustworthy Runtime Verification via Bisimulation (Extended Experience Report)}

\author{Ryan G. Scott}
\affiliation{%
  \institution{Galois, Inc.}
  \country{USA}
  \authoremail{rscott@galois.com}
}

\author{Ivan Perez}
\affiliation{%
  \institution{KBR @ NASA Ames Research Center}
  \country{USA}
  \authoremail{ivan.perezdominguez@nasa.gov}
}

\author{Alwyn E. Goodloe}
\affiliation{%
  \institution{NASA Langley Research Center}
  \country{USA}
  \authoremail{a.goodloe@nasa.gov}
}

\author{Mike Dodds}
\affiliation{%
  \institution{Galois, Inc.}
  \country{USA}
  \authoremail{miked@galois.com}
}

\author{Robert Dockins}
\authornote{Author's paper contributions were made while working at Galois, Inc.}
\affiliation{%
  \institution{Amazon}
  \country{USA}
  \authoremail{rdoc@amazon.com}
}



\begin{abstract}
When runtime verification is used to monitor safety-critical systems, it is
essential that monitoring code behaves correctly.
The \copilot{} runtime verification framework pursues this goal by
automatically generating C monitor programs from a high-level DSL embedded in
Haskell.
In safety-critical domains, every piece of deployed code must be
accompanied by an assurance argument that is convincing to human auditors.
However, it is difficult for auditors to determine with confidence that a
compiled monitor cannot crash and implements the behavior required by the
\copilot{} semantics.

In this paper we describe \verifier{}, which runs alongside the \copilot{}
compiler, generating a proof of correctness for the compiled output.
The proof establishes that a given \copilot{} monitor and its compiled form
produce equivalent outputs on equivalent inputs, and that they either crash in
identical circumstances or cannot crash.
The proof takes the form of a bisimulation broken down into a set of
verification conditions.
We leverage two pieces of SMT-backed technology: the \crucible{} symbolic
execution library for LLVM and the \whatFour{} solver interface library.
Our results demonstrate that dramatically increased compiler assurance can be
achieved at moderate cost by building on existing tools.
This paves the way to our ultimate goal of generating formal assurance
arguments that are convincing to human auditors.

\end{abstract}

\maketitle

\section{Introduction}
\label{sec:introduction}

Safety-critical cyber-physical systems (CPSs) are subject to strict regulation
to ensure public safety.
Historically, such systems are constructed using conservative
requirements-driven practices, yielding predictable systems that are
amenable to verification by testing~\citep{ARP4754A,DO178C}.
There is increasingly a desire to use off-the-shelf components and employ
techniques like machine learning to build autonomous systems, but these cannot
be assured using traditional approaches~\citep{NRC14,
AFE87,cofer2020,cofer2021}.

Runtime verification (RV)~\citep{GoodloePike10,FalconeHR13} addresses this
problem by monitoring a system under observation and responding to property
violations during the mission.
For example, an RV system might monitor engine heat levels, aircraft location
within an authorized airspace, or autopilot changes between flight modes.
While not static formal verification, RV provides a significant improvement in
assurance for systems over testing alone.

\copilot{}~\citep{pike2010copilot,pike2013copilot,perez2020copilot} is a
language and toolchain for writing RV monitors.
Monitors are written in a high-level, stream-based domain-specific language
(DSL) that is embedded in Haskell.
\copilot{} is equipped with a compiler that generates C code, which can then be
linked against other application code for use in production.
Among others, \copilot{} is used at NASA to write both programs and runtime
monitors for experimental vehicles (e.g., drones, rovers).

When used in safety-critical applications, \copilot{} monitors also form part
of the safety case that justifies the system's mission readiness.
As a result, monitors must be trustworthy, and trust must be based on
systematic and rigorous evidence that can be audited.
Specifically, \copilot{} monitors must be trustworthy in two regards.
First, the monitor must be \emph{validated}, meaning that the monitor enforces
the higher-level properties that were intended.
Second, the C implementation of the monitor must be \emph{verified}, meaning
that the compiled code correctly realizes the expected semantics of the
monitor.
The problem of \emph{validation} is about establishing properties of the
monitor program itself, and the problem of \emph{verification} is about the
correctness of the \copilot{} compiler.

Several mechanisms can be used to validate and verify the monitors.
In recent years, formal methods have been increasingly accepted as a means of
evidence in domains such as civil aerospace.
For example, DO-333~\citep{DO333} gives official guidance on the use of formal
methods in certification practice.
Specifically, with regards to verification, there are generally three main
approaches that can be used to formally verify the correctness of the generated
code:
\begin{enumerate}
\item Verifying the entire compiler end to end, such as in
CompCert~\citep{compcert} or \velus~\citep{Bourke2017, Bourke2019, Bourke2021}.
\item \emph{Proof-carrying code}~\citep{1997:necula:pcc}, where the compiler
can produce Hoare-style annotations in the compiled code, allowing the proofs
to be checked independently.
\item \emph{Translation validation}~\citep{pnueli1998translation}, where the
code is compiled first, and a proof of correctness for the particular program
is produced afterwards.
\end{enumerate}

This paper presents \verifier{}, which addresses the problem of verification
using the translation validation approach.\footnote{For validation, \copilot{}
supports reasoning about monitor specs using theorem
provers~\citep{laurent2015assuring,goodloe2016challenges}.}
\verifier{} proves that the \copilot{} monitor and the C program are
\emph{extensionally equal} according to each language's executable semantics:
given a stream of inputs, the monitor and its compiled form produce equivalent
outputs.
We also prove that either (1) both are memory safe, i.e. they cannot crash, or
(2) that they crash exactly on equivalent inputs (the tool supports either
mode).

To prove the equivalence between a \copilot{} specification and the generated C
program, \verifier{} establishes a bisimulation between them.
More precisely, at each program point, \verifier{} proves that the state of the
\copilot{} monitor corresponds to the state of the C program.
Our verifier is designed to be push-button: given a \copilot{} monitor as
input, the verifier automates the steps needed to construct a proof.

This problem of compiler correctness is familiar from research projects such as
CompCert and CakeML~\citep{cakeml}, but these have tended to be multi-year
multi-person efforts.
This level of effort was not available to us as an industry project operating
under budget constraints.
Our objective was more pragmatic: to maximize our confidence in the \copilot{}
compiler using off-the-shelf formal methods tools and libraries.
We make several design choices that reduce the cost of building \verifier{}
but, in some cases, reduce the power of our results (see
\Cref{sec:design-tradeoffs}).
We judge these to be low-cost tradeoffs relative to the large increase in
confidence achieved by successfully completing the \verifier{} project in just
under one year.
Overall, our results have been heartening: through careful design, \verifier{}
shows that it is possible to apply formal assurance to a DSL compiler with
modest levels of effort.

Our ultimate goal, of which this project is the first step, is to automatically
generate formal evidence that is convincing to human auditors.
By providing a proof of equivalence between the monitor implementation and the
original specification as supporting evidence, we aim to build a certification
argument that strongly relies on formal methods, rather than being based mostly
on software engineering practices.

The rest of the paper is structured as follows.
In \Cref{sec:copilot-overview}, we describe the \copilot{} language and
illustrate the problem that \verifier{} addresses.
In \Cref{sec:verifier-architecture}, we give a high-level description of
\verifier{}, and we illustrate our work by showing bugs in \copilot{} that
\verifier{} helped identify.
\Cref{sec:bisimulation} discusses the bisimulation relation between \copilot{}
and C programs.
\Cref{sec:assurance-cases} presents an example of evidence that the verifier
produces for assurance cases.
\Cref{sec:case-study} showcases case studies involving verification of
aerospace-related monitors.
In \Cref{sec:design-tradeoffs} we elaborate on the design trade-offs made in
\verifier{}.
\Cref{sec:implementation} describes the implementation of the verifier.
We close the paper with a discussion on related work (\Cref{sec:related-work})
and conclusions (\Cref{sec:conclusions}).

\section{\copilot{} overview}
\label{sec:copilot-overview}

\copilot{} is an embedded domain-specific language (EDSL)
implemented as a Haskell library.
This section provides a brief overview of the \copilot{} language and the
aspects of its semantics relevant to our verification task.
Readers interested in further details should consult \linebreak \citet{perez2020copilot}.
Readers familiar with \copilot{} can skip to \Cref{sec:challengecopilot} for a discussion
of the challenge of verifying \copilot{} compilation.
Note that we generally refer to programs written in the \copilot{} DSL as
\emph{monitors}, but they are also often called \emph{specs} or
\emph{specifications}, reflecting their role in runtime verification.

\subsection{\copilot{} streams}

The main programming abstraction in \copilot{} is the \emph{stream}, which
represents a discrete, infinite sequence of values over time.
Values include integers, floating-point numbers, booleans, and
compound types like structs and arrays.
\copilot{} programs are evaluated one step
at a time, either at regular intervals or as new data becomes available.
The time between each value in a stream is abstract
and application-dependent; it can be thought
of as a constant corresponding to a unit of real time.

To interact with the system being monitored, \copilot{} programs may reference
\emph{external} streams, which generally represent values
obtained from the system being monitored (e.g., sensor data).
From these external streams, other intermediate streams of data are computed,
eventually resulting in boolean streams that capture the
conditions being monitored.
At execution time, an external handler function is called whenever a \emph{trigger
stream} evaluates to true, which allows the execution environment to take action.

\Cref{fig:thermostat_ex} shows a complete \copilot{} program implementing a
simple thermostat.
The program monitors an external stream (\lstinline|temp|) representing data
from a temperature probe, firing a \lstinline|trigger| whenever the temperature
(\lstinline|avgTemp|) is sufficiently above or below a fixed setpoint, each
captured by a boolean stream in the \lstinline|spec|.
To guard against noisy data, the probe input data is smoothed by computing a
sliding window average of the last 5 samples.

This program demonstrates how to construct streams of constant values using
\lstinline|constant|, as well as how to cast from a stream of
\lstinline|Word8|s to a stream of \lstinline|Float|s using
\lstinline|unsafeCast|.\footnote{This cast is safe, since there is no information loss
in this direction. Casting in the other direction could lose
information. \copilot{} uses type classes to enforce which type conversions
should be treated as ``safe'' or ``unsafe''.}
The thermostat program also demonstrates some operations on streams whose
implementations differ from the Haskell functions of the same name.
Operations such as multiplication \lstinline|(*)|, subtraction \lstinline|(-)|,
and comparison \lstinline|(>)| work pointwise over the elements of a stream.
The \lstinline|(++)| operator prepends a list containing a fixed number of
samples to the front of a stream.
Invoking \lstinline|sum n s| will compute a stream where the $i$th element of
the stream consists of the sum of the elements in \lstinline|s| from indices
$i$ through $i \; +$ (\lstinline|n| $- \;1)$.

\begin{figure*}[h]
\begin{center}
\begin{tabular}{c}
\begin{lstlisting}[language=Haskell]
-- External temperature as a byte, from -50C to 100C
temp :: Stream Word8
temp = extern "temperature" Nothing

-- Calculate temperature in Celsius
ctemp :: Stream Float
ctemp =
  (unsafeCast temp*constant (150.0/255.0)) - constant 50.0

-- Width of the sliding window
window :: Int
window = 5

-- Compute sliding average of last 5 temps
avgTemp :: Stream Float
avgTemp = (sum window (replicate window 19.5 ++ ctemp))
        / fromIntegral window

spec :: Spec
spec = do
  trigger "heaton"  (avgTemp < 18.0) [arg avgTemp]
  trigger "heatoff" (avgTemp > 21.0) [arg avgTemp]
\end{lstlisting}
\end{tabular}
\end{center}
\caption{A simple thermostat example.}
\label{fig:thermostat_ex}
\end{figure*}

\subsection{Reifying \copilot{} specifications}

The first step when working with any \copilot{} monitor (whether to simulate,
prove properties, or generate code) is to \emph{reify} the monitor into an
intermediate representation called \copilot{} Core.
In addition to unfolding helper definitions to produce the intermediate syntax,
the reification step checks that the streams verified are \emph{well-formed}.
Well-formedness is a syntactic check that ensures that no computations violate
temporal causality (i.e., they only depend on ``past'' values) and that they
require only a finite history to compute.
These properties ensure that monitors can be implemented using a step-by-step
strategy and in constant memory.

The reified version of the thermostat example is shown in
\Cref{fig:thermostat_ex_reified}.
Here, it becomes clear how the sliding window calculation is unfolded into a
stream computation involving the \lstinline|drop| operator to sum up the 5 most
recent values of the stream named \lstinline|s0|.
Given a number \lstinline|n|, \lstinline|drop n| drops the first \lstinline|n|
elements from a stream, where the stream must have at least \lstinline|n|
elements of history available.
In our example, we prepend 5 constant values to the stream \lstinline|s0| so
there is always enough history, even at the beginning of execution.
The \lstinline|cast| operator is the reified counterpart to
\lstinline|unsafeCast| in the stream language.

\begin{figure*}
\begin{center}
\begin{tabular}{c}
\begin{lstlisting}[language=Haskell]
(Float) s0 =
    [19.5,19.5,19.5,19.5,19.5]
      ++ (((cast) Ext_temperature * 0.5882353) - 50.0)
trigger "heaton" =
   ((((((s0 + drop 1 s0) + drop 2 s0) + drop 3 s0)
     + drop 4 s0) / 5.0) < 18.0)
   [arg (((((s0 + drop 1 s0) + drop 2 s0) + drop 3 s0)
     + drop 4 s0) / 5.0)]
trigger "heatoff" =
   ((((((s0 + drop 1 s0) + drop 2 s0) + drop 3 s0)
     + drop 4 s0) / 5.0) > 21.0)
   [arg (((((s0 + drop 1 s0) + drop 2 s0) + drop 3 s0)
     + drop 4 s0) / 5.0)]
\end{lstlisting}
\end{tabular}
\end{center}
\caption{The reified version of \lstinline|spec| in \Cref{fig:thermostat_ex}.}
\label{fig:thermostat_ex_reified}
\end{figure*}

\subsection{Compiling to C code}

The \copilotC{} library takes a reified monitor and generates C code.
The generated code is not a standalone program and is intended to be used part
of a larger system.
To ease integration, the code has no external dependencies and targets the C99
subset of the C language.

There is a straightforward translation from each possible stream value to a C
value.
For instance, an \lstinline|Int32| in a stream program would be translated to
an \lstinline|int32_t| in C, and similarly for other scalar types.
Compound stream values such as structs and arrays are translated to C structs
and arrays with corresponding struct field names and array lengths.

While \copilot{} streams are conceptually infinite sequences,
\copilot{}-generated C programs only use a finite amount of memory.
Each stream is translated to a ring buffer, implemented as
an array with fixed length equal to the history needed to compute the
stream.
Each buffer has an associated index that tracks the current position in the
array, which is incremented as time advances.
Note that all streams advance time in lock-step.

\Cref{fig:thermostat_ex_c} shows the generated C code for the thermostat
example.  The \lstinline|s0| stream is translated to C as a ring buffer with 5
elements, just large enough to store the initial 5 temperature values.
This ring buffer will be used to compute the current values for the various
streams appearing in the trigger definitions, and it will be updated based on
the current value of the external \lstinline|temperature| stream on each tick.
A more complicated monitor may have additional ring buffers of various lengths,
each of which will be updated as necessary.

The \lstinline|step| function advances the state of the program by a single
tick or time step.
The following diagram shows how \lstinline|step| computes subsequent
temperatures from the previous time step and inserts into the appropriate
position in the buffer, tracked by the index value \lstinline|s0_idx|.
This diagram assumes that the inputs from the external \lstinline|temperature|
stream begin $[150,95,... ]$.
One additional temperature value is computed at each time step, which is marked
in this diagram in red italic font.

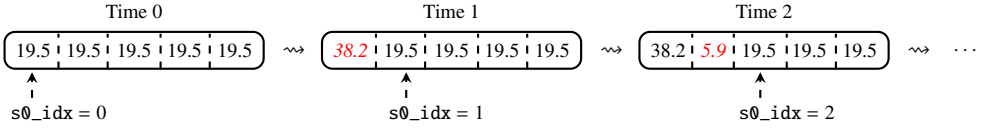
\begin{center}
 \begin{tikzpicture}[
             > = stealth, 
             shorten > = 1pt, 
             auto,
             node distance = 2.5cm, 
             thick 
         ]
 \begin{scope}[scale=0.75, transform shape]
   \node[nodeStyle] (cZero) {19.5 \; 19.5 \; 19.5 \; 19.5 \; 19.5};
   \draw[dashed,transform canvas={xshift=-1.00cm}] (cZero.south)--(cZero.north);
   \draw[dashed,transform canvas={xshift=-0.35cm}] (cZero.south)--(cZero.north);
   \draw[dashed,transform canvas={xshift= 0.35cm}] (cZero.south)--(cZero.north);
   \draw[dashed,transform canvas={xshift= 1.00cm}] (cZero.south)--(cZero.north);

   \node[nodeStyle] (cOne) [right=1.0cm of cZero] {\textcolor{red}{\it 38.2} \; 19.5 \; 19.5 \; 19.5 \; 19.5};
   \draw[dashed,transform canvas={xshift=-1.00cm}] (cOne.south)--(cOne.north);
   \draw[dashed,transform canvas={xshift=-0.35cm}] (cOne.south)--(cOne.north);
   \draw[dashed,transform canvas={xshift= 0.35cm}] (cOne.south)--(cOne.north);
   \draw[dashed,transform canvas={xshift= 1.00cm}] (cOne.south)--(cOne.north);

   \node[nodeStyle] (cTwo) [right=1.0cm of cOne] {38.2 \; \textcolor{red}{\it 5.9} \; 19.5 \; 19.5 \; 19.5};
   \draw[dashed,transform canvas={xshift=-0.95cm}] (cTwo.south)--(cTwo.north);
   \draw[dashed,transform canvas={xshift=-0.40cm}] (cTwo.south)--(cTwo.north);
   \draw[dashed,transform canvas={xshift= 0.30cm}] (cTwo.south)--(cTwo.north);
   \draw[dashed,transform canvas={xshift= 0.95cm}] (cTwo.south)--(cTwo.north);

   \node[] (cThree) [right=1.0cm of cTwo] {$\dots$};


   \node[] (cIndexZero)  [below=0.5cm of cZero,transform canvas={xshift=-1.00cm}] {\lstinline{s0_idx} $= 0$};
   \node[] (cIndexOne)   [below=0.5cm of cOne, transform canvas={xshift=-0.20cm}] {\lstinline{s0_idx} $= 1$};
   \node[] (cIndexTwo)   [below=0.5cm of cTwo, transform canvas={xshift= 0.30cm}] {\lstinline{s0_idx} $= 2$};

   \path[edgeStyle,dashed,transform canvas={xshift=-1.35cm}] (cIndexZero) edge node {} (cZero);
   \path[edgeStyle,dashed,transform canvas={xshift=-0.60cm}] (cIndexOne) edge node {} (cOne);
   \path[edgeStyle,dashed,transform canvas={xshift=-0.05cm}] (cIndexTwo) edge node {} (cTwo);

   \node[] (cSighOne) [below=0.5cm of cZero] {};


   \node[] (streamZero)  [above=0.1cm of cZero] {Time 0};
   \node[] (streamOne)   [above=0.1cm of cOne]  {Time 1};
   \node[] (streamTwo)   [above=0.1cm of cTwo]  {Time 2};


  \node[] (squig01)  [right=0.20cm of cZero] {$\rightsquigarrow$};
  \node[] (squig02)  [right=0.20cm of cOne] {$\rightsquigarrow$};
  \node[] (squig03)  [right=0.20cm of cTwo] {$\rightsquigarrow$};

 \end{scope}
\end{tikzpicture}
\end{center}

Aside from updating the ring buffer, the \lstinline|step| function is
responsible for checking if any of the monitored conditions have been met and,
if so, firing the corresponding triggers.
The trigger functions \lstinline|heaton| and \lstinline|heatoff| are only given
forward declarations; they are meant to be defined by the application that
links against the \copilot{} monitor.

\begin{figure*}

\begin{center}
\begin{tabular}{c}
\begin{lstlisting}[language=C]
extern uint8_t temperature;
void heaton(float arg0);
void heatoff(float arg0);

static float s0[5] = {19.5f, 19.5f, 19.5f, 19.5f, 19.5f};
static size_t s0_idx = 0;

float s0_get(size_t x) {
  return s0[(s0_idx + x) % 5];
}

float avgTemp(void) {
 return (s0_get(0) + s0_get(1) + s0_get(2)
         + s0_get(3) + s0_get(4)) / 5.0f;
}

void step(void) {
  if (avgTemp() < 18.0f) heaton(avgTemp());
  if (avgTemp() > 21.0f) heatoff(avgTemp());
  s0[s0_idx] = (temperature * 0.5882353f) - 50.0f;
  s0_idx = (s0_idx + 1) % 5;
}
\end{lstlisting}
\end{tabular}
\end{center}

\caption{C code generated for \Cref{fig:thermostat_ex}.
The code has been cleaned up for presentation purposes only.}
\label{fig:thermostat_ex_c}
\end{figure*}

\subsection{The challenge of verifying \copilot{} compilation}
\label{sec:challengecopilot}

We have seen how the thermostat example has been translated from a stream-based
program (\Cref{fig:thermostat_ex}) to a C program (\Cref{fig:thermostat_ex_c}),
but can we be sure that they actually behave in identical ways?
With careful evaluation, one can see that the values in the \lstinline|s0|
buffer, as computed by \lstinline|s0_get(0)|, \lstinline|s0_get(1)|, $\dots$,
and \lstinline|s0_get(4)| at a given time step $n$, will match the $n$th,
$(n+1)$th, $\dots$, and $(n+4)$th elements of the \lstinline|avgTemp| stream.
As a result, the \lstinline|heaton| and \lstinline|heatoff| trigger functions
in C will be called if and only if the corresponding trigger streams evaluate
to true, and the arguments to the trigger functions will always equal the value
of the \lstinline|avgTemp| stream at that time step.
In this sense, the two programs exhibit the same behavior.
Most \copilot{} monitors in the wild (e.g., \Cref{sec:case-study}) are
significantly more involved than this example, however, and establishing a
correspondence between the monitor and the generated C code is substantially
harder.

\section{Verifier overview}
\label{sec:verifier-architecture}

\verifier{} runs alongside the \copilotC{} compiler and formally verifies that
the results are correct.
Specifically, it proves that the semantics of the input \copilot{} monitor
correspond to the semantics of the generated C program.
In most cases, the verifier requires no input annotations beyond what is
supplied to the compiler (we discuss exceptions in
\Cref{sec:automation-limits}).
The implementation of the verifier can be found at \citet{zenodoArtifact}.

\definecolor{verifiedGreen}{RGB}{196,255,205}
\definecolor{falsifiedRed}{RGB}{255,204,196}

\begin{figure*}
 \begin{center}
 \begin{tikzpicture}[
             > = stealth, 
             shorten > = 1pt, 
             auto,
             node distance = 2.5cm, 
             thick 
         ]
 \begin{scope}[scale=0.65, transform shape]
  \node[nodeStyle] (streamProgram) {\copilot{} monitor};
   \node[nodeStyle] (cProgram) [below=1cm of streamProgram] {\copilot{} C program};
   \node[nodeStyle] (llvmBitcode) [below=1cm of cProgram] {LLVM bitcode};
   \node[nodeStyle] (crucible) [below=1cm of llvmBitcode] {\crucible{}};
   \node[nodeStyle] (streamSemantics) [right=2cm of cProgram] {\copilot{} semantics {\bf (1)}};
   \node[nodeStyle] (cSemantics) [below=1cm of streamSemantics] {C/LLVM semantics {\bf (2)}};
   \node[nodeStyle] (bisimulation) [right=1cm of streamSemantics] {Bisimulation prover {\bf (3)}};
   \node[nodeStyle,fill=verifiedGreen] (allGoalsVerified) [right=1cm of bisimulation] {All goals verified};
   \node[nodeStyle,fill=falsifiedRed] (falsifiedGoals) [below=1cm of allGoalsVerified] {Falsified goals};
   \node[draw,dotted,fit=(streamSemantics) (cSemantics)] {\footnotesize \whatFour{}};
   \node[nodeStyle] (assuranceCase) [above=1cm of allGoalsVerified] {Assurance case};

   \path[edgeStyle] (streamProgram) edge[right] node {\copilotC} (cProgram);
   \path[edgeStyle] (streamProgram.east) edge[bend left] node {\copilotTheorem} (streamSemantics);
   \path[edgeStyle] (cProgram) edge[right] node {Clang} (llvmBitcode);
   \path[edgeStyle] (llvmBitcode) edge[right] node {\crucibleLLVM} (crucible);
   \path[edgeStyle] (crucible.east) edge[bend right,below right] node {Symbolic simulation} (cSemantics);
   \path[edgeStyle] (streamSemantics) edge node {} (bisimulation);
   \path[edgeStyle] (cSemantics) edge[bend right] node {} (bisimulation.south west);
   \path[edgeStyle] (bisimulation) edge[loop below ] node {SMT solver} ();
   \path[edgeStyle] (bisimulation) edge node {} (allGoalsVerified);
   \path[edgeStyle] (bisimulation.south east) edge[bend right] node {} (falsifiedGoals);
   \path[edgeStyle] (allGoalsVerified) edge node {} (assuranceCase);
 \end{scope}
 \end{tikzpicture}

 \caption{The architecture of \verifier.}
 \label{fig:architecture}

 \end{center}
\end{figure*}

The architecture of \verifier{} is depicted in \Cref{fig:architecture}.
At its heart is a comparison between the input \copilot{} program, which
computes over streams {\bf (1)}, and the compiled C program, which represents
streams in real memory {\bf (2)}.
To enable this comparison, both the input and compiled program are represented
by a collection of formulas in \whatFour{}, an SMT interface
library~\citep{what4}.
The relationship between the two semantics are then established by the
bisimulation prover {\bf (3)}.
\verifier{} reports back to the user whether it solved all of the generated
proof goals successfully.
Specifically, the verifier either produces a successful result, produces a
failing result, or, in exceptional cases, diverges when solving proof goals
(\Cref{sec:smt-solver-performance}).
If it succeeds, the verifier produces an assurance argument that the proof is
valid, as explained in \Cref{sec:assurance-cases}.
If it fails, the verifier identifies which goals were falsified.

Intuitively, programs at both the \copilot{} and C levels are translated to
transition systems corresponding to the program control-flow graph.
The formulas that are generated in \whatFour{} encode the effect of a single
transition on the state---either streams or memory.
The generated transition systems are intentionally very similar in structure,
so the main task of the bisimulation proof is to demonstrate that the
states stay in correspondence (see
Section~\ref{sec:bisimulation}).

The translation from \copilot{} programs to \whatFour{} is performed by
\copilotTheorem{}, which we extended as part of this work.
\copilotTheorem{} uses \citet{perez2020copilot} as the source of truth for the
semantics of \copilot{} stream programs.
Because \copilot{} is a functional language, it is relatively simple to encode
the semantics of each program step in an SMT style.
For instance, \copilot's integer arithmetic operations involving streams
translate straightforwardly to SMT-Lib's fixed-size bitvector operations, so
\copilotTheorem{} would translate the expression \lstinline|x + 42 :: Stream Int32|
into an SMT formula resembling \lstinline|(bvadd x (_ bv42 32)))|.
There are similarly direct translations for all other stream operations with
the exception of floating-point operations, a special case that we discuss
further in \Cref{sec:floating-point}.

It is much more complex to faithfully capture the semantics of a C program.
To do this, we lean on a pre-existing tool, Galois's \crucible~\citep{crucible}
symbolic simulation library.
Specifically, \verifier{} compiles the C program to LLVM bitcode and uses
\crucible's LLVM backend to simulate it.
The result is a collection of \whatFour{} formulas that precisely represent the
LLVM program's semantics.
\crucible{} provides an accurate model of LLVM, intended for industry formal
methods applications.
For example, \crucible{} has previously been applied in tools used to verify
industry cryptographic
libraries~\citep{saw,continuous-verification-s2n,Boston2021}.

Continuing the earlier example, when \crucible{} simulates \lstinline|x + 42|
(where \lstinline|x| is an \lstinline|int32_t|) as C code, it returns two SMT
formulas.
The first SMT formula is \lstinline|(bvadd x (_ bv42 32)))|, which represents
the result of the overall expression.
The second SMT formula is \linebreak
\lstinline|(bvslt x (_ bv2147483616 32)))|, which must be satisfiable for the
overall expression to be free of undefined behavior.
This formula encodes the fact that the C code performs signed integer addition,
and signed integer overflow is undefined behavior in C---hence, \lstinline|x|
must be less than $2^{31} - 32 = 2147483616$.
See \Cref{sec:partial-operations} for more discussion on how \verifier{}
handles C operations that are not well-defined on all inputs.

\paragraph{Bugs found in \copilot}
\label{sec:copilot-bugs}

We developed \verifier{} by testing it against a variety of examples used in
\copilot's test suite, which uncovered a number of bugs in \copilot{} itself.
These issues include memory unsafety~\citep{copilotIssue:238,copilotIssue:386},
incorrect C code
generation~\citep{copilotIssue:275,copilotIssue:276,copilotIssue:314,copilotIssue:373},
and mismatches between \copilot{} and C
semantics~\citep{copilotIssue:278,copilotIssue:263}.
To illustrate one particular bug in more detail, \verifier{} was able to detect
that the translation of the \lstinline|signum| function in \copilotC{} was
inconsistent with the \copilot{} semantics~\citep{copilotIssue:278}.
Specifically, the former would claim that \lstinline|signum 0 = 1|, but the
latter would state that \lstinline|signum 0 = 0|.
This bug, as well as all the other previously mentioned issues, have been fixed
upstream.

\section{Bisimulation proof structure}
\label{sec:bisimulation}

The core theorem we wish to prove is \emph{extensional equality}.
That is, the \copilot{} program and its compiled C representation behave
identically in the input-output behavior observable in the execution context.
Since only trigger functions can be observed in \copilot{}, we opt for the
following formal description.
A \copilot{} program $P$ and its compiled form $Q$ are extensionally equal if
for any arbitrary input stream, the following holds at every time step:
\begin{itemize}
  \item The same set of trigger functions are called in $P$ and $Q$ with the
    same arguments.
  \item $P$ has crashed iff $C$ has crashed.
\end{itemize}
Note that \verifier{} checks for the absence of undefined behavior in the
generated C code (\Cref{sec:partial-operations}).
Therefore, we do not need to consider non-termination in our setting, since the
generated C code contains no loops, no recursion, and no potentially unsafe or
diverging function calls.

Proving extensional equality between arbitrary programs is difficult, but the
\copilot{} program and the compiled program are intentionally very similar in
their structure.
Consider again the \copilot{} thermostat program in ~\Cref{fig:thermostat_ex}
and the resulting C code in ~\Cref{fig:thermostat_ex_c} (we will use this
as a running example in this section).
Let us assume that the C program's first $n$ inputs correspond to the first $n$
values of the external stream inputs.
Intuitively, after $n$ calls to the \lstinline|step| function in the C program,
the state of the ring buffers should be equal to the value of the corresponding
stream expressions at index $n$.
Moreover, the trigger functions in the C program should be called from the
\lstinline|step| function at the same times when the corresponding stream
expressions evaluate to true.

We can view a \copilot{} stream program and its generated C program as labeled
transition systems (LTSes).
To prove correctness, \verifier{} constructs a bisimulation relation between
the two systems.
Intuitively, the proof shows that the two systems start in corresponding
states, and that every transition in one system has a transition to a
corresponding state in the other system.
This has the effect of proving extensional equality because it shows that
trigger functions are called in corresponding states, and that the two systems
transition to crashing states at the same time.

The bisimulation constructed by \verifier{} is a \emph{strong}
bisimulation~\citep{1981:park:strong-bisimulation,1990:milner:strong-weak-bisimulation}
because every transition in one system corresponds exactly to a transition in
the other system.
Note that this is not the only way we could have proven a bisimulation.
Alternatively, we could have constructed a
\emph{weak}~\citep{1990:milner:strong-weak-bisimulation} or
\emph{branching}~\citep{1996:van-glabbeek:branching-bisimulation,1996:basten:branching-bisimulation}
bisimulation, where some internal states in one system are abstracted away in
the other system.
In our very restricted setting, it is more convenient to use a strong
bisimulation, as one time step in \copilot{} naturally corresponds one
invocation of the \lstinline|step| function in C.

Using a strong bisimulation is somewhat unusual from a compiler verification
perspective, where the system representing compiled code is usually an
\emph{abstraction} of the specification's system, i.e., where multiple states
in the compiled code can correspond to a single state in the specification.
We instead use a strong bisimulation because we are allowed to set up
\verifier{} in a way where the states and transitions naturally line up, so it
is more convenient to build an LTS for the compiled C code that combines all of
the internal actions of a \lstinline|step| function into a single transition.

To be more precise, \verifier{} does not prove the overall bisimulation
property.
Instead, it proves a set of per-transition properties whose conjunction implies
a bisimulation.
Verifying the bisimulation would take us outside the logical fragment that can
be easily reasoned about in SMT solvers.


\subsection{Programs as transition systems}

Formally, an LTS consists of a set of states, a set of labels, and a set of
labeled transitions between pairs of states.
%
Consider again the thermostat example
(\Cref{fig:thermostat_ex,fig:thermostat_ex_c}).
Assume that the inputs from the external
\lstinline|temperature| stream begin $[150,95,... ]$,
which are roughly $38.2^\circ$ and $5.9^\circ$ in Celsius.
%
%
Note that the triggers \lstinline|heaton| and \lstinline|heatoff| will fire
if the sliding average of the previous five temperatures dips below $18^\circ$
or if it exceeds $21^\circ$, respectively.


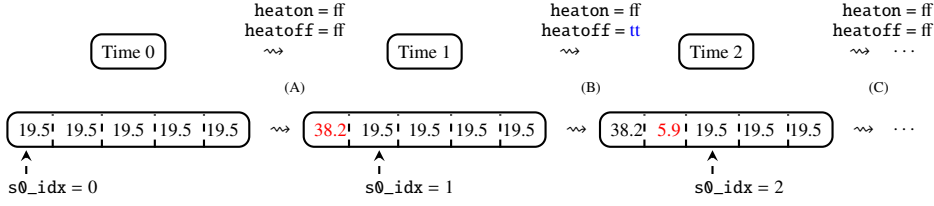
\begin{figure}
 \begin{center}
 \vspace*{0.6cm}
 \begin{tikzpicture}[
             > = stealth, 
             shorten > = 1pt, 
             auto,
             node distance = 2.5cm, 
             thick 
         ]
 \begin{scope}[scale=0.7, transform shape]
   \node[nodeStyle] (cZero) {19.5 \; 19.5 \; 19.5 \; 19.5 \; 19.5};
   \draw[dashed,transform canvas={xshift=-1.00cm}] (cZero.south)--(cZero.north);
   \draw[dashed,transform canvas={xshift=-0.35cm}] (cZero.south)--(cZero.north);
   \draw[dashed,transform canvas={xshift= 0.35cm}] (cZero.south)--(cZero.north);
   \draw[dashed,transform canvas={xshift= 1.00cm}] (cZero.south)--(cZero.north);

   \node[nodeStyle] (cOne) [right=1.0cm of cZero] {\textcolor{red}{38.2} \; 19.5 \; 19.5 \; 19.5 \; 19.5};
   \draw[dashed,transform canvas={xshift=-1.00cm}] (cOne.south)--(cOne.north);
   \draw[dashed,transform canvas={xshift=-0.35cm}] (cOne.south)--(cOne.north);
   \draw[dashed,transform canvas={xshift= 0.35cm}] (cOne.south)--(cOne.north);
   \draw[dashed,transform canvas={xshift= 1.00cm}] (cOne.south)--(cOne.north);

   \node[nodeStyle] (cTwo) [right=1.0cm of cOne] {38.2 \; \textcolor{red}{5.9} \; 19.5 \; 19.5 \; 19.5};
   \draw[dashed,transform canvas={xshift=-0.95cm}] (cTwo.south)--(cTwo.north);
   \draw[dashed,transform canvas={xshift=-0.40cm}] (cTwo.south)--(cTwo.north);
   \draw[dashed,transform canvas={xshift= 0.30cm}] (cTwo.south)--(cTwo.north);
   \draw[dashed,transform canvas={xshift= 0.95cm}] (cTwo.south)--(cTwo.north);

   \node[] (cThree) [right=1.0cm of cTwo] {$\dots$};


  \node[] (squig01)  [right=0.25cm of cZero] {$\rightsquigarrow$};
  \node[] (squig02)  [right=0.25cm of cOne] {$\rightsquigarrow$};
  \node[] (squig03)  [right=0.25cm of cTwo] {$\rightsquigarrow$};

   \node[] (cIndexZero)  [below=0.5cm of cZero,transform canvas={xshift=-1.00cm}] {\lstinline{s0_idx} $= 0$};
   \node[] (cIndexOne)   [below=0.5cm of cOne, transform canvas={xshift=-0.20cm}] {\lstinline{s0_idx} $= 1$};
   \node[] (cIndexTwo)   [below=0.5cm of cTwo, transform canvas={xshift= 0.30cm}] {\lstinline{s0_idx} $= 2$};

   \path[edgeStyle,dashed,transform canvas={xshift=-1.35cm}] (cIndexZero) edge node {} (cZero);
   \path[edgeStyle,dashed,transform canvas={xshift=-0.60cm}] (cIndexOne) edge node {} (cOne);
   \path[edgeStyle,dashed,transform canvas={xshift=-0.05cm}] (cIndexTwo) edge node {} (cTwo);

   \node[] (cSighZero) [above=1cm of cZero] {};
   \node[] (cSighOne) [below=0.5cm of cZero] {};


   \node[nodeStyle] (streamZero)  [above=0.8cm of cZero] {Time 0};
   \node[nodeStyle] (streamOne)   [above=0.8cm of cOne]  {Time 1};
   \node[nodeStyle] (streamTwo)   [above=0.8cm of cTwo]  {Time 2};
   \node[]          (streamThree) [right=2.50cm of streamTwo] {$\dots$};


  \node[] (squig01)  [right=1.71cm of streamZero] {$\rightsquigarrow$};
  \node[] (squig02)  [right=1.71cm of streamOne] {$\rightsquigarrow$};
  \node[] (squig03)  [right=1.68cm of streamTwo] {$\rightsquigarrow$};

  \node[] (transA)  [below=0.1cm of streamZero,transform canvas={xshift=+2.2cm}] {\footnotesize (A)};
  \node[] (transB)  [below=0.1cm of streamOne,transform canvas={xshift=+2.2cm}] {\footnotesize (B)};
  \node[] (transC)  [below=0.1cm of streamTwo,transform canvas={xshift=+2.15cm}] {\footnotesize (C)};

  \node[align=right] (trigger0)  [above=0.6cm of transA,transform canvas={xshift=+2.2cm}] {\lstinline{heaton} = ff \\[-0.1cm] \lstinline{heatoff} = ff};
  \node[align=right] (trigger1)  [above=0.6cm of transB,transform canvas={xshift=+2.2cm}] {\lstinline{heaton} = ff \\[-0.1cm] \lstinline{heatoff} = \textcolor{blue}{tt}};
  \node[align=right] (trigger2)  [above=0.6cm of transC,transform canvas={xshift=+2.2cm}] {\lstinline{heaton} = ff \\[-0.1cm] \lstinline{heatoff} = ff};

 \end{scope}
\end{tikzpicture}

 \caption{Two labelled transition systems (LTSes). The upper LTS represents the stream
 program in \Cref{fig:thermostat_ex}. The lower LTS represents the C
 program in \Cref{fig:thermostat_ex_c}. Transition labels are shared
 between both LTSes. Transition B fires the \lstinline|heatoff|
 trigger.}

 \label{fig:thermostat_ltses}

\end{center}
\end{figure}

\copilot{} is a stream processing language based on functional programming
ideas.
As a result, it has no explicit state in its semantics, beyond the (immutable)
input streams and the current time step.
Instead, the value of stream expressions are calculated from the input stream
at the current time step and its prefix.
An example sequence of transitions for the thermostat stream program with
concrete inputs is pictured in the upper diagram of
\Cref{fig:thermostat_ltses}.
%

In a C program, each state consists of the global memory that contains the ring
buffers and their current indices.
The transition relation is defined by the generated \lstinline|step| function.
An LTS for the thermostat C program is pictured
in the lower diagram of \Cref{fig:thermostat_ltses}.
This program's global memory only tracks one buffer, \lstinline|s0|, which
holds the five most recent temperatures.
The global memory also tracks \lstinline|s0_idx|, the current index into the
buffer.
In each transition, a newly sampled value from \lstinline|temperature| is
placed in the position that \lstinline|s0_idx| points to, after which the index is
incremented.

The same set of labels is used for both the stream and C LTSes.
Each label is associated with observable events required for the transition to
occur.
In the stream spec, the events correspond to trigger streams evaluating to
true.
In C, the events correspond to the \lstinline|step| function
invoking the corresponding trigger functions.

In \Cref{fig:thermostat_ltses}, each transition records whether the
\lstinline|heaton| and \lstinline|heatoff| triggers fired.
The \lstinline|heatoff| trigger fires in transition (B), as adding $38.2^\circ$
makes the sliding average temperature $23.2^\circ$, which exceeds the upper
bounds that \lstinline|avgTemp| checks for.
In transition (C), however, \lstinline|heatoff| no longer fires, as adding
$5.9^\circ$ lowers the average temperature to $20.5^\circ$.

\subsection{Correspondence relation}
\label{sec:correspondence-relation}

We now define a \emph{correspondence relation} between stream and C program
states.
In \copilot{}, each stream has a finite window of past values that can be
accessed at any point in the program.
Consider a \copilot{} program containing a stream $s$ with a window value $k$.
Let \lstinline|buf| be the ring buffer in the C program that corresponds to
$s$, and let \lstinline|idx| be the current index into \lstinline|buf|.
A stream program state is related to a C program state by the correspondence
relation if and only if the value of $s$ at index $n+i$ is equal to the value
of \lstinline|buf| at index (\lstinline|idx| $+ \; i) \; \text{mod} \; k$,
where $i$ ranges from $0$ to $k-1$.
We lift this to sets of streams in the obvious way.

The thermostat example has a single stream definition \lstinline|s0| that
retains \lstinline|window = 5| previous values.
At time step 0, the \lstinline|s0| stream's first five elements are
\lstinline|19.5| due to the use of \lstinline|replicate window 19.5| in its
definition.
In C, the values of \lstinline|s0_get(i)| (as \lstinline|i| ranges from
\lstinline|0| to \lstinline|4|) are also \lstinline|19.5|, as these are the
initial values of the \lstinline|s0| buffer before running \lstinline|step|.
Therefore, the two programs are in correspondence at time step 0.
We can also intuit that the two programs correspond at subsequent time steps.
For instance, at time step 1, the \lstinline|s0| stream would produce
\lstinline|38.2| as its fifth value, which would match the value of
\lstinline|s0_get(4)| after a single invocation of the \lstinline|step|
function on the C side.

\subsection{Proving the bisimulation}

The goal of \verifier{} is to demonstrate that the correspondence
relation is a bisimulation.
That is, for every pair of states $(s, c)$ in the correspondence relation, if
$s$ transitions to $s'$ in the stream program with label $\alpha$, then there
must exist a C program state $c'$ such that $c$ transitions to $c'$ with label
$\alpha$.
The converse must also be true: if $c$ transitions to $c'$, then there must
exist a stream program state $s'$ such that $s$ transitions to $s'$ with label
$\alpha$.

It is straightforward to identify which transitions in each LTS correspond to
each other, as the $n$th time step in the stream program corresponds to the
$n$th invocation of \lstinline|step| in the C program.
The main challenge, then, is to demonstrate that the stream values equal the
ring buffer values at each time step.
The verifier does this by proving three properties about the correspondence
relation:

\begin{enumerate}

\item The initial states of the programs are in the correspondence relation.

\item For each pair of states $(s, c)$ in the correspondence relation, there
exist a stream program state $s'$, a C program state $c'$, and a label $\alpha$
such that $s$ transitions to $s'$ with label $\alpha$, $c$ transitions to $c'$
with label $\alpha$, and $(s', c')$ is in the correspondence relation.

\item For each label $\alpha$ used in the transition relations, the triggers
for $\alpha$ fire in corresponding ways in both programs.

\end{enumerate}

Note that the definition of a bisimulation has two directions: one direction in
which the stream program state $s'$ is universally quantified and the C program
state $c'$ is existentially quantified, and another direction in which the
order of quantification is reversed.
Property (2), on the other hand, checks both directions simultaneously.
This is done for practical reasons, as it is always clear what the
existentially quantified states $s'$ and $c'$ are after each transition.
The verifier combines both directions into a single step with the same label
$\alpha$.

\verifier{} reduces each of these properties to SMT formulas.
The proof principle of bisimulation itself is not amenable to SMT, as it falls
outside of the first-order theories that SMT solvers understand.
Likewise, the semantics of \copilot{} and C could potentially be reduced to
SMT, but it would be impractical to do so.
Instead, we reduce the individual proof obligations above to lower-level
logical statements that can be represented as SMT queries.

\subsubsection{Initial state correspondence}

The first proof obligation that the verifier must discharge is that the initial
states of the two programs correspond.
This is tantamount to taking the initial values in each ring buffer and proving
that they are equal to the first $k$ values of the corresponding stream at time
step 0, where $k$ is the length of the ring buffer.
Due to the restrictions that \copilot{} places on programs, these first $k$
values (e.g., the \lstinline|19.5| values prepended to \lstinline|s0| in
\Cref{fig:thermostat_ex_reified}) must be concrete and cannot depend on
external inputs.
As a result, this step is simple to translate to SMT queries and only requires
evaluation of concrete values.

{\bf Thermostat example:} We demonstrated initial state correspondence for the
thermostat example in \Cref{sec:correspondence-relation}.
The proof is tantamount to showing that the values of the
\lstinline|s0| stream and ring buffer all equal \lstinline|19.5| at time step
0, a total of five proof goals.

\subsubsection{Transition correspondence}

Most of the proof effort consists of demonstrating that the correspondence
relation is preserved by transitions.
In this phase of the proof, we begin with completely symbolic program states at
an arbitrary time value $n$.\footnote{In
\Cref{sec:invariants-for-partial-operations}, we explain how to constrain
symbolic program states by adding invariants.}
As a result, we must create fresh symbolic values for each stream definition
and its corresponding ring buffer.

More precisely, let $s$ range over the stream definitions, where each $s$ is
required to retain $k$ previous values.
Let \lstinline|buf| be the ring buffer in the C program that corresponds to
$s$, and let \lstinline|idx| be the current index into \lstinline|buf|.
For each $i$ ranging from $0$ to $k - 1$, we create a symbolic value and assume
it is equal to both the value of $s$ at index $n+i$ and the value of
\lstinline|buf| at index (\lstinline|idx| $+ \; i) \; \text{mod} \; k$.
We then advance the symbolic state of the stream and C programs once.
Then, for each $j$ ranging from $1$ to $k$, we read the value of $s$ at index
$n+j$ and check that it is equal to the value of \lstinline|buf| at index
(\lstinline|idx| $+ \; j) \; \text{mod} \; k$ under the previous assumptions.

Advancing the symbolic state of the \copilot{} stream program is a matter of
evaluating each stream expression at the next time step.
For the C program, we advance the symbolic state by invoking the \crucible{}
symbolic simulator to run the \lstinline|step| function, which updates the
memory used in the program.
In addition to generating proof goals about bisimulation equivalence
conditions, \crucible{} can also generate side conditions that relate to the
memory safety of the program.
For instance, each memory access into a ring buffer could potentially have
out-of-bounds indexing.
If \copilotC{} generates C code correctly, all such accesses should be within
bounds, but this must be checked as a part of the query submitted to the SMT
solver.
The simulator also generates side conditions related to C operations with
undefined behavior---see \Cref{sec:partial-operations} for details.

%
{\bf Thermostat example:}
Ten goals involve checking symbolic values for equality, with two goals
discharged for each element that the \lstinline|s0| stream retains.
Thirty goals involve checking if the array indexes in \lstinline|s0_get| and
\lstinline|step| are within bounds.
Four goals involve checking if the trigger functions fire in equivalent ways,
which is described in \Cref{sec:proving-triggers}.
In total, the verifier discharges 44 goals.

\subsubsection{Triggers}
\label{sec:proving-triggers}


The proof must establish that, for each trigger function in the generated C
program, the trigger function is called if and only if the corresponding
trigger stream in the stream program evaluates to true.
In a real setting, the application linked against a \copilot{} monitor would
implement the trigger functions. However, we are verifying the generated
monitor code in isolation, so we do not have implementations of the trigger
functions available. Instead, the verifier creates stub implementations for
each trigger function that captures the arguments and the path condition under
which it was called.
If no path in the C program reaches a trigger, then a false path condition is
generated.
After symbolic simulation finishes, the captured arguments and path condition
are asserted to be equivalent to the corresponding trigger stream and arguments
from the stream program.

{\bf Thermostat example:}
Four proof goals arise from checking for equivalent behavior for the
\lstinline|heaton| and \lstinline|heatoff| triggers.
Two goals check that each trigger is fired in both programs during a
transition.
Two goals check that the arguments passed to each trigger correspond.

\subsubsection{Partial operations}
\label{sec:partial-operations}

Another subtlety of the transition relation step of the proof is how to handle
partial \copilot{} operations. These range from division, which can fail if the
second argument is zero, to signed integer arithmetic, which can overflow.
If a partial operation is used on an input for which it is not
defined, it can result in undefined behavior in the generated C code.

One way to handle partial operations is to take an uncompromising approach: if
\crucible{} detects undefined behavior when simulating the generated C code, it
aborts and causes the proof to fail. This ensures that \copilot{}
specifications do not misbehave, but, if a spec \emph{does} misbehave, the
verifier will simply fail and not reveal anything about the rest of the spec.

Another way to handle partial operations is to prove that a \copilot{} spec is
``crash-equivalent'' to its corresponding C program. That is, if the C program
invokes a partial operation on undefined inputs, the verifier checks that the
corresponding operation in the \copilot{} spec is invoked on the same inputs.
\verifier{} supports both the uncompromising approach and the crash-equivalence
approaches as user-configurable options.

To check for crash-equivalence during symbolic simulation, the verifier will
analyze any invocation of a partial operation in the stream program.
For each invocation, \verifier{} will generate a side condition that this
operation will only be invoked on well defined inputs.
During the transition step of the proof, the verifier will assume these side
conditions before starting symbolic simulation.
Therefore, if the simulator generates any side conditions due to partial
operations in the C program, they should be dischargeable using the
corresponding side conditions from the stream program.


\section{Assurance cases}
\label{sec:assurance-cases}

\verifier{} is the first step in a longer project: integrating formal methods
into safety-critical deployment practices.
Our ultimate goal is for \copilot{} code to be deployed with a safety case
largely constructed by automated means.
Existing standards such as DO-333 provide guidance on how formal methods
evidence should be handled.
However, there are few examples of successful formal methods deployments in
safety-critical practice.
See~\citep{10.1007/978-3-319-57288-8_29,10.1007/978-3-319-06200-6_1} for a
discussion of the issues involved in certification and formal methods.

To use \copilot{} in safety-critical systems, the evidence provided by our
tools must be understood and accepted by human auditors.
This is challenging because \verifier{} performs most of its reasoning through
complicated SMT queries.
These queries are difficult to analyze manually without some way of connecting
them back to higher-level requirements.

We address this challenge by giving \verifier{} a setting that, when enabled,
displays each \crucible{} proof goal generated during a successful run of the
verifier.
Each proof goal has an accompanying high-level description of what it
demonstrates, such as a symbolic stream value being equal to its corresponding
C ring buffer value.
Each proof goal also has an associated \whatFour{} formula and SMT query
representing the goal.
With this information, each portion of a proof can be broken down into lower-
and lower-level requirements until eventually reaching SMT, with a chain of
evidence linking each intermediate step.

\tikzstyle{goal}=[
    draw = black,
    thick,
    minimum size = 4mm,
    rectangle,
    inner sep = 2mm,
    align=center
]
\tikzstyle{strategy}=[
    draw = black,
    thick,
    minimum size = 4mm,
    rectangle,
    inner sep = 2mm,
    align=center,
    trapezium,
    trapezium left angle=60,
    trapezium right angle=120
]

\newsavebox\goalSix
\newsavebox\goalSeven
\newsavebox\goalEight
\newsavebox\goalNine

\begin{lrbox}{\goalSix}
\begin{lstlisting}
let v121 =
      bvUrem
        (bvSum cs0_idx@89:bv 0x4:[64])
        0x5:[64]
    v122 =
      bvSum
        (bvMul 0x4:[64] v121)
    ...
\end{lstlisting}
\end{lrbox}

\begin{lrbox}{\goalSeven}
\begin{lstlisting}
let v94 =
      bvSum
        (bvMul 0x4:[64] cs0_idx@89:bv)
    v100 =
      bvUrem
        (bvSum cs0_idx@89:bv 0x1:[64])
        0x5:[64]
    v1874 =
      bvUrem
        (bvSum v100 0x1:[64])
          0x5:[64]
    ...
\end{lstlisting}
\end{lrbox}

\begin{lrbox}{\goalEight}
\begin{lstlisting}
(declare-fun s0_idx () (_ BitVec 64))
(define-fun x!0 () (_ BitVec 64)
  (bvadd s0_idx (_ bv4 64)))
(define-fun x!1 () (_ BitVec 64)
  (bvurem x!0 (_ bv5 64)))
(define-fun x!2 () (_ BitVec 64)
  (bvmul (_ bv4 64) x!1))
...
\end{lstlisting}
\end{lrbox}

\begin{lrbox}{\goalNine}
\begin{lstlisting}
(declare-fun s0_idx () (_ BitVec 64))
(define-fun x!0 () (_ BitVec 64)
  (bvmul (_ bv4 64) s0_idx))
(define-fun x!1 () (_ BitVec 64)
  (bvadd s0_idx (_ bv1 64)))
(define-fun x!2 () (_ BitVec 64)
  (bvurem x!1 (_ bv5 64)))
...
\end{lstlisting}
\end{lrbox}

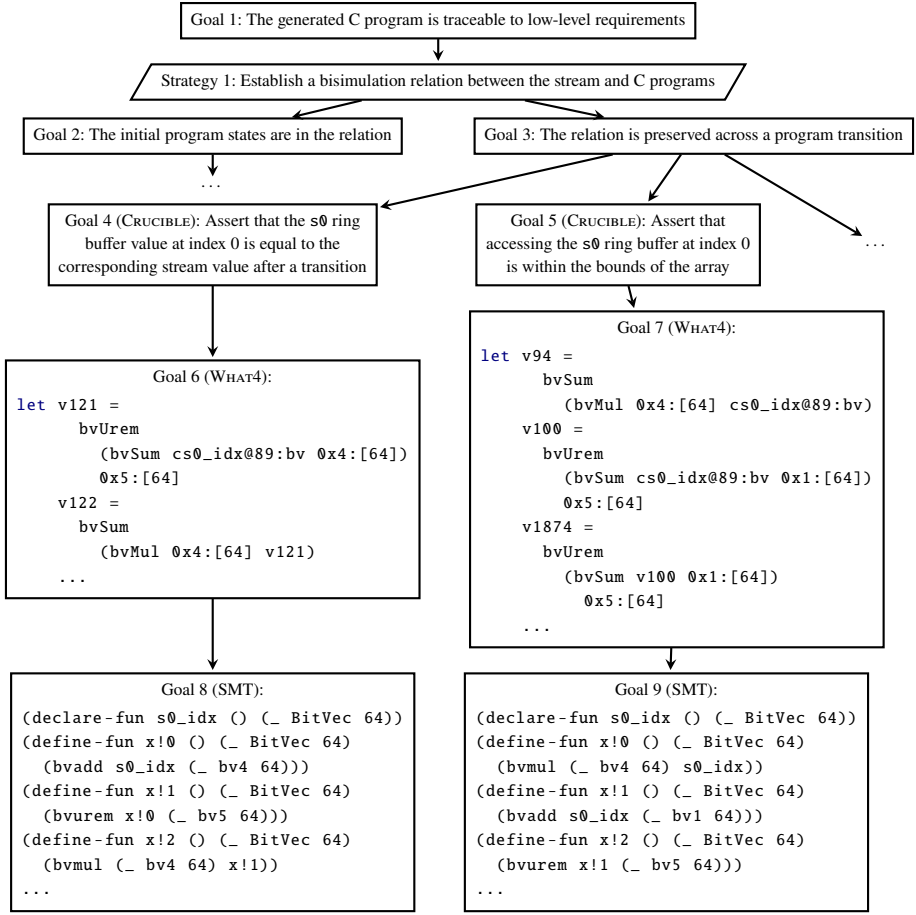
\begin{figure*}
 \begin{center}
 \begin{tikzpicture}[
             > = stealth, 
             shorten > = 1pt, 
             auto,
             node distance = 2.5cm, 
             thick 
         ]
 \begin{scope}[scale=0.65, transform shape]

  \node[goal] (goal1) {Goal 1: The generated C program is traceable to low-level requirements};
  \node[strategy] (strategy1) [below=0.5cm of goal1] {Strategy 1: Establish a bisimulation relation between the stream and C programs};
  \node[goal] (initGoal) [below left=0.5cm of strategy1] {Goal 2: The initial program states are in the relation};
  \node (initGoalDots) [below=0.5cm of initGoal] {\dots};
  \node[goal] (transGoal) [below right=0.5cm of strategy1]
    {Goal 3: The relation is preserved across a program transition};
  \node[goal] (crucibleGoal1) [below=1cm of initGoal]
    {Goal 4 (\crucible): Assert that the \lstinline|s0| ring \\ buffer value at index 0 is equal to the \\ corresponding stream value after a transition};
  \node[goal] (crucibleGoal2) [right=2cm of crucibleGoal1]
    {Goal 5 (\crucible): Assert that \\ accessing the \lstinline|s0| ring buffer at index 0 \\ is within the bounds of the array};
  \node (crucibleGoalDots) [right=2cm of crucibleGoal2] {\dots};
  \node[goal] (what4Goal1) [below=1.5cm of crucibleGoal1]
    {Goal 6 (\whatFour): \\
     \usebox\goalSix
    };
  \node[goal] (what4Goal2) [right=1cm of what4Goal1]
    {Goal 7 (\whatFour): \\
     \usebox\goalSeven
    };
  \node[goal] (smtGoal1) [below=1.5cm of what4Goal1]
    {Goal 8 (SMT): \\
     \usebox\goalEight
    };
  \node[goal] (smtGoal2) [right=1cm of smtGoal1]
    {Goal 9 (SMT): \\
     \usebox\goalNine
    };

  \path[edgeStyle] (goal1) edge node {} (strategy1);
  \path[edgeStyle] (strategy1) edge node {} (initGoal);
  \path[edgeStyle] (strategy1) edge node {} (transGoal);
  \path[edgeStyle] (initGoal) edge node {} (initGoalDots);
  \path[edgeStyle] (transGoal) edge node {} (crucibleGoal1);
  \path[edgeStyle] (transGoal) edge node {} (crucibleGoal2);
  \path[edgeStyle] (transGoal) edge node {} (crucibleGoalDots);
  \path[edgeStyle] (crucibleGoal1) edge node {} (what4Goal1);
  \path[edgeStyle] (crucibleGoal2) edge node {} (what4Goal2);
  \path[edgeStyle] (what4Goal1) edge node {} (smtGoal1);
  \path[edgeStyle] (what4Goal2) edge node {} (smtGoal2);

 \end{scope}
 \end{tikzpicture}

 \caption{A sketch of a GSN diagram presenting a \verifier{} assurance case for
          the thermostat programs from \Cref{fig:thermostat_ex,fig:thermostat_ex_c}. For
          brevity's sake, only a subset of goals are shown.}
 \label{fig:gsn-example}

 \end{center}
\end{figure*}

Using the evidence that \verifier{} produces, one can construct a complete
assurance case that is more amenable to certification.
\Cref{fig:gsn-example} shows a sketch of what an assurance case would look like
for the thermostat programs from \Cref{fig:thermostat_ex,fig:thermostat_ex_c}.
The evidence is captured using \emph{Goal Structuring Notation} (GSN)
\citep{gsn}, which presents each goal in a \verifier{} safety case in a way that
emphasizes the relationship between high-level parent goals and lower-level
child goals.
Aside from proof goals, a GSN diagram can also be used to document assumptions
that the verifier makes during verification, such as those discussed in
\Cref{sec:design-tradeoffs}.

The process of making \verifier's evidence be acceptable for auditors is
ongoing work.
We plan to work with auditors to iteratively improve formats and explanations
towards the goal of providing convincing evidence in a suitable format.

\section{Case studies}
\label{sec:case-study}

In this section, we report on the use of \verifier{} to verify monitors of
specific aerospace-related algorithms and safety monitors used in experimental
flights.

\subsection{Sense and avoid}

We turn our attention to using \verifier{} to verify a \copilot{}
implementation of an algorithm used in an unmanned aircraft system (UAS) to
\emph{sense and avoid} other aircraft.
More precisely, the \cite{sense-and-avoid} defines the concept of sense and avoid
as ``the capability of a UAS to remain well clear from and avoid collisions
with other airborne traffic''.
\citet{upchurch2014analysis} provides a rigorous well-clear boundary model
where any UAS inside the boundary are considered to be in well-clear violation.
\copilot{} is particularly well-suited to monitoring well-clear violations,
which makes \citet{upchurch2014analysis}'s algorithm an interesting case study
for how well \verifier{} can handle a real-world aerospace-related use case.

The first step in the verification process is to implement the well-clear
violation checking algorithm inside of \copilot.
For the most part, this is a straightforward exercise, as all of the
mathematical operations that the algorithm uses directly correspond to
floating-point operations that \copilot{} supports.
We review some highlights of the \copilot{} implementation below, but readers
interested in further details should consult the full implementation
\footnote{\url{https://github.com/Copilot-Language/copilot-verifier/blob/cea9f8e924fe8f50cdbffd15ea7e00d7f73f1d8a/copilot-verifier/examples/Copilot/Verifier/Examples/ShouldPass/WCV.hs}},
as well as the original algorithm's description in
\citet{upchurch2014analysis}.

The algorithm heavily relies on computing horizontal and vertical components in
3-D space.
Vertical components (comprising the z-axis) are described as points (i.e.,
altitude values), and horizontal components (comprising the x- and y-axis) are
described as 2-D vectors within 3-D space.
For convenience, we define a \lstinline|Vect2| type alias in \copilot{} to
refer to vector values, which consist of a pair of two
\lstinline|Double|-valued streams:

\begin{center}
\begin{tabular}[t]{c}
\begin{lstlisting}[language=Haskell]
type Vect2 = (Stream Double, Stream Double)
\end{lstlisting}
\end{tabular}
\end{center}

We also define common vector operations such as negation, dot product,
squaring, and computing vector lengths and determinants.
For instance, vector dot product is defined as:

\begin{center}
\begin{tabular}[t]{c}
\begin{lstlisting}[language=Haskell]
(|*|) :: Vect2 -> Vect2 -> Stream Double
(|*|) (x1, y1) (x2, y2) = (x1 * x2) + (y1 * y2)
\end{lstlisting}
\end{tabular}
\end{center}

Well-clear detection relies on knowing the relative velocities and positions
between the \emph{ownship} aircraft and the \emph{intruder} aircraft.
By convention, we represent the horizontal relative velocity and position as
vectors (i.e., as \lstinline|Vect2|s) named \lstinline|v| and \lstinline|s|,
and we represent the vertical relative velocity and position as points (i.e.,
as streams of \lstinline|Double|s) named \lstinline|vz| and \lstinline|sz|.
All computations within the algorithm are defined in terms of these quantities.

To detect if there is a well-clear violation, the algorithm needs a
\emph{(horizontal) time variable}, which is a function that maps the horizontal
relative position and velocity into a real number.
One such time variable is the \emph{time of horizontal closest point of
approach}, which we abbreviate as \lstinline|tcpa| in the implementation.
Based on the horizontal relative position and velocity, \lstinline|tcpa s v|
computes a positive number if the aircraft are horizontally converging, a
negative number if they are horizontally diverging, and the number zero
otherwise:

\begin{center}
\begin{tabular}[t]{c}
\begin{lstlisting}[language=Haskell]
tcpa :: Vect2 -> Vect2 -> Stream Double
tcpa s v@(vx, vy) =
  if vx ~= 0 && vy ~= 0 then 0 else -(s |*| v)/(sq v)
\end{lstlisting}
\end{tabular}
\end{center}

Note that \lstinline|sq| computes the square of a vector, and \lstinline|(~=)|
checks if two vectors are equal plus or minus a small error value.
The \lstinline|tcpa| function is one possible time variable, but it is not the
only one; there are alternative time variables \citep{upchurch2014analysis},
which we also implement.

In addition to \lstinline|tcpa|, which describes horizontal closeness, we also
compute the \emph{time to co-altitude} (\lstinline|tcoa|), which describes if
the aircraft are vertically converging or diverging:

\begin{center}
\begin{tabular}[t]{c}
\begin{lstlisting}[language=Haskell]
tcoa :: Stream Double -> Stream Double -> Stream Double
tcoa sz vz = if (sz * vz) < 0
               then (-sz) / vz
               else -1
\end{lstlisting}
\end{tabular}
\end{center}

Related to \lstinline|tcpa| is the \emph{distance} at time of closest point
approach, or \lstinline|dcpa|:

\begin{center}
\begin{tabular}[t]{c}
\begin{lstlisting}[language=Haskell]
dcpa :: Vect2 -> Vect2 -> Stream Double
dcpa s@(sx, sy) v@(vx, vy) =
  norm (sx + (tcpa s v) * vx, sy + (tcpa s v) * vy)
\end{lstlisting}
\end{tabular}
\end{center}

With these functions defined, we now have everything we need to define the
notion of a well-clear violation (\lstinline|wcv|).
Such a violation occurs when there is both a horizontal and vertical violation.
Observe that the \lstinline|wcv| function is parameterized not just by the
positions and velocities, but also the horizontal time variable
\lstinline|tvar|:

\begin{center}
\begin{tabular}[t]{c}
\begin{lstlisting}[language=Haskell]
wcv ::
  (Vect2 -> Vect2 -> Stream Double) ->
  Vect2 -> Stream Double ->
  Vect2 -> Stream Double ->
  Stream Bool
wcv tvar s sz v vz =
  horizontalWCV tvar s v && verticalWCV sz vz
\end{lstlisting}
\end{tabular}
\end{center}

A horizontal violation occurs when the relative horizontal position falls below
a certain threshold \lstinline|dthr|, or if the distance at time of closest
point of approach is less than \lstinline|dthr| and the time variable falls
below a certain threshold \lstinline|tthr|:

\begin{center}
\begin{tabular}[t]{c}
\begin{lstlisting}[language=Haskell]
horizontalWCV ::
  (Vect2 -> Vect2 -> Stream Double) ->
  Vect2 -> Vect2 -> Stream Bool
horizontalWCV tvar s v =
  (norm s <= dthr) ||
  (((dcpa s v) <= dthr) &&
    (0 <= (tvar s v)) && ((tvar s v) <= tthr))
\end{lstlisting}
\end{tabular}
\end{center}

A vertical violation occurs when the relative vertical position is less than a
given threshold \lstinline|zthr|, or if the time to co-altitude is less then a
given threshold \lstinline|tcoathr|:

\begin{center}
\begin{tabular}[t]{c}
\begin{lstlisting}[language=Haskell]
verticalWCV ::
  Stream Double -> Stream Double -> Stream Bool
verticalWCV sz vz =
  ((abs $ sz) <= zthr) ||
  (0 <= (tcoa sz vz) && (tcoa sz vz) <= tcoathr)
\end{lstlisting}
\end{tabular}
\end{center}

The exact values of the threshold values depend on the specifics of the UAS in
question.
We define the thresholds as external streams.
Note that, although this allows us to vary the thresholds over time, we expect
the thresholds to remain constant in this specific application:

\begin{center}
\begin{tabular}[t]{c}
\begin{lstlisting}[language=Haskell]
dthr, tthr, zthr, tcoathr :: Stream Double
dthr    = extern "dthr" Nothing
tthr    = extern "tthr" Nothing
zthr    = extern "zthr" Nothing
tcoathr = extern "tcoathr" Nothing
\end{lstlisting}
\end{tabular}
\end{center}

Similarly, we sample the relative positions and velocities from the
application:

\begin{center}
\begin{tabular}[t]{c}
\begin{lstlisting}[language=Haskell]
vx, vy, vz :: Stream Double
vx = extern "relative_velocity_x" Nothing
vy = extern "relative_velocity_y" Nothing
vz = extern "relative_velocity_z" Nothing

sx, sy, sz :: Stream Double
sx = extern "relative_position_x" Nothing
sy = extern "relative_position_y" Nothing
sz = extern "relative_position_z" Nothing

-- Relative velocity and position as 2D vectors.
v, s :: Vect2
v = (vx, vy)
s = (sx, sy)
\end{lstlisting}
\end{tabular}
\end{center}

Finally, we can define the overall \copilot{} specification:

\begin{center}
\begin{tabular}[t]{c}
\begin{lstlisting}[language=Haskell]
spec :: Spec
spec =
  trigger "well_clear_violation" (wcv tcpa s sz v vz) []
\end{lstlisting}
\end{tabular}
\end{center}

One interesting property of this \copilot{} spec is that all of the basic
streams are external ones.
As such, when compiling this spec to C code, the resulting code does not need
to use any ring buffers to store stream elements; all stream values are
represented as C \lstinline|extern|s:

\begin{center}
\begin{tabular}[t]{c}
\begin{lstlisting}[language=Haskell]
extern double relative_position_x;
extern double relative_position_y;
extern double relative_velocity_x;
extern double relative_velocity_y;
extern double relative_position_z;
extern double relative_velocity_z;
extern double dthr;
extern double tthr;
extern double zthr;
extern double tcoathr;
\end{lstlisting}
\end{tabular}
\end{center}

Using entirely external \copilot{} streams does not pose an issue for
\verifier.
In fact, this property of the spec arguably makes it \emph{easier} to verify,
as it makes checking the initial state of the bisimulation relation extremely
simple.
\verifier{} represents both the values of external \copilot{} streams and the
values of C \lstinline|extern|s using the same fresh, symbolic variables, which
mean that they are trivially equal to each other.

The only interesting part of verifying this spec is proving the transition step
of the bisimulation relation.
Because \verifier{} treats all floating-point operations as uninterpreted
functions (see ~\Cref{sec:floating-point}), it does not perform any deep
reasoning about partial floating-point functions such as division.
As such, it does not emit any proof goals related to memory safety in the
transition step.
There is only one interesting proof goal emitting during this step: proving
that the \lstinline|well_clear_violation| trigger fires in the \copilot{} spec
if and only if the corresponding C handling function is called.
The proof goal is quite large due to the multitude of floating-point
operations, but the verifier is nevertheless able to solve the proof goal in
less than one second.

\subsection{Monitoring safe mode changes}

\verifier{} has been used to verify the correctness of monitors flown in drones
at NASA Langley Research Center (LaRC).
We worked with a team at LaRC to capture flight software properties in
\copilot{} and produce a NASA Core Flight System
(cFS)~\cite{2005:wilmot:cfs,github:cfs} application that monitored the vehicle
during flight.
That application monitored the status of the vehicle based on information
published by other components, and alerted of potentially dangerous mode
changes before they occurred (e.g., transitioning to autopilot without a GPS
signal).

To help justify that the \copilot{} monitors could be flown safely in the same
vehicle, we included (1) the proof of correctness of the generated C code,
obtained using \verifier{}, (2) evidence that the time and memory consumed by
the monitors was bounded, obtained using randomized tests and tools that measure
memory usage, and (3) evidence that the monitors would not crash or raise
exceptions (also obtained using randomized tests).

\section{Design tradeoffs}
\label{sec:design-tradeoffs}

When designing \verifier{}, our goal was to create a verification
tool that achieves high assurance at modest cost.
This section explains the tradeoffs involved in achieving this.

\subsection{Trusted computing base (TCB)}

We trust the underlying C toolchain.
\verifier{} relies on Clang to compile C into LLVM bitcode, which becomes the
basis for producing the semantics of the C file.
Bugs in Clang may affect the soundness of the verifier.
We consider this risk mitigated by the fact that Clang is a well-tested
compiler and that \copilotC{} targets a well-understood subset of C, reducing
the likelihood of triggering compiler bugs.
The \copilot{} developers have experimented using CompCert to verify the
compiled binary code~\citep{compcert,goodloe2016challenges}, which would remove
this portion of the TCB.
At the time, CompCert did not target many of the architectures that the
\copilot{} users utilize, but CompCert has since added new backends, including
RISC-V and AArch64.
We plan on revisiting the use of CompCert as future work.

When \verifier{} runs, it uses the Clang version available to the user.
In principle, the meaning of the C program could change if the program is
compiled using different flags, a different Clang version, or a different C
compiler altogether.
We leave it to users to ensure that their choice of compiler does not alter the
behavior of the program in a meaningful way.

We trust that \crucible's LLVM backend faithfully encodes the semantics of LLVM
bitcode. Errors in this part of \crucible{} could affect the soundness of the
verifier.
To justify this trust, we note that \crux{}~\citep{crux}, a verification tool
also based on \crucible, has been tested on a large number of C verification
problems from the SV-COMP verification competition~\citep{cruxSVComp}.

We also trust that the \copilotTheorem{} library accurately encodes the
semantics of \copilot{} stream programs. At present, we do not have a robust
way to test these semantics beyond careful engineering and manual comparison
with the \copilot{} interpreter.

Because of the way we model trigger functions,
we make a number of implicit assumptions about how the
implementations of those functions must behave.
In particular, we assume that
trigger functions do not modify any memory under the control of the \copilot{}
program, including its ring buffers and stack.
We also assume that the trigger
functions are memory-safe and do not perform any undefined behavior.
Responsibility for enforcing these assumptions lies with the user who supplies
definitions for the trigger functions.

\subsection{C compiler optimizations}
\label{sec:c-compiler-optimizations}

\verifier{} assumes that all streams in a \copilot{} spec have corresponding
static, global arrays in the generated LLVM bitcode.
This assumption simplifies the work done by \verifier{},
as it can associate each stream with a distinct, top-level
array.

While this assumption is safe to make when the generated C code is compiled
with low optimization settings, it is \emph{not} safe at higher settings.
For instance, consider a stream containing a single value, which \verifier{}
assumes to be translated to an array of length 1.
At \lstinline|-O1| or higher, Clang replaces length-1 static arrays with scalar
values, which breaks the verifier's assumptions.

The verifier mitigates these issues by always invoking Clang with
\lstinline|-O0|.
This makes the generated LLVM bitcode more predictable at the expense of only
verifying code at lower levels of optimization.
When code is compiled to be deployed on the target system, and after a
successful run of \verifier{}, users can invoke Clang with higher optimization
settings and have an expectation that the resulting code will match the
original specification, provided that Clang correctly optimizes the code.
We consider that the task of verifying the correctness of Clang to be out of
the scope of our work.

\subsection{Limits of automation}
\label{sec:automation-limits}

\verifier{} aims to be a push-button solution, and, in our experiments, we
successfully used it to verify all the examples in the \copilot{} test suite
(\Cref{sec:copilot-bugs}) and two case studies (\Cref{sec:case-study}).
However, in exceptional circumstances, \verifier{} may not be able to prove
that a generated program is equivalent to its specification.
That is, the verifier is \emph{sound, but not complete}.
When this occurs, users may be able to alter the original spec to make it
amenable to verification.
We provide examples in the rest of this section.

\subsubsection{Floating-point support}
\label{sec:floating-point}

\copilot{} provides a variety of floating-point operations, including
transcendental functions and other primitive functions.
There is limited SMT solver support for floating-point values, however, and
even less support for robustly handling special functions.
Nevertheless, we wish to have \emph{some} level of support, as many \copilot{}
monitors make essential use of floating-point operations.
For instance, detecting if an unmanned aircraft system is \emph{well
clear}~\citep{upchurch2014analysis} uses the square-root function to compute the
lengths of vectors.

\verifier{} treats floating-point operations as uninterpreted functions,
leaving the semantics of the operations abstract. This approach is sound, as
the verifier need only demonstrate that a \copilot{} program applies the same
floating-point operations as the corresponding C program, and in the same
order. The downside to this technique is that reasoning about floating-point
operations is somewhat fragile. The verifier relies on the Clang not optimizing
the operations to the point where they differ from the stream program's
operations. As an example, consider \lstinline|ctemp| from
\Cref{fig:thermostat_ex}:

\begin{lstlisting}[language=Haskell]
ctemp :: Stream Float
ctemp =
  (unsafeCast temp * constant (150.0 / 255.0))
    - constant 50.0
\end{lstlisting}

A subtlety of this definition is that the division operation occurs before the
numeric value is lifted into a stream with the \lstinline|constant| function.
As a result, the reified stream in \Cref{fig:thermostat_ex_reified} contains
\lstinline|0.5882353|, the result of the division. If, instead, we instead
express it as the result of dividing two stream values:

\begin{lstlisting}[language=Haskell]
ctemp =
  (unsafeCast temp) * (constant 150.0 / constant 255.0))
    - constant 50.0
\end{lstlisting}

\noindent then the reified \copilot{} spec would \emph{not} evaluate the
result of dividing the two streams to \lstinline|0.5882353|. On the other
hand, the generated C code would contain \lstinline|150.f / 255.0f|, and C
compilers will perform constant folding on this, even on low optimization
levels. As a result, the stream semantics would contain an uninterpreted
division function but the C semantics would not, leading an SMT solver to
conclude that they may not be equivalent.

We take some measures to mitigate this issue, such as invoking C compilers on
low optimization settings (see \Cref{sec:c-compiler-optimizations}) and
avoiding the use of the \lstinline[language={}, emphstyle={}]{--fast-math}
flag.
Under these settings, C compilers, and Clang in particular, are more reluctant
to rearrange floating-point code, which results in more programs successfully
verifying.
Nevertheless, the example above shows that this is not foolproof, and users may
need to rearrange code to make the operations align in just the right way.

\subsubsection{Invariants for partial operations}
\label{sec:invariants-for-partial-operations}

As detailed in \Cref{sec:partial-operations}, the verifier has a mode for
aborting the proof early if it finds a partial operation applied to
undefined inputs. In this mode, the verifier does not try to infer
invariants needed to make operations well defined.
For example:
\begin{center}
\begin{tabular}[t]{c}
\begin{lstlisting}[language=Haskell]
stream :: Stream Int32
stream = extern "stream" Nothing

spec :: Spec
spec = trigger "streamAdd"
  ((stream + 1) == 2) [arg stream]
\end{lstlisting}
\end{tabular}
\end{center}

When abort-early mode for partial operations is enabled, this example fails to
verify, as \lstinline|stream + 1| could result in signed integer overflow if a
value in \lstinline|stream| is equal to the maximum value of an
\lstinline|Int32|. It is possible, however, for applications that use this
monitor to maintain the invariant that the values in \lstinline|stream| are
always less than the maximum \lstinline|Int32| size. To do so, the user must
communicate this invariant by declaring a property in the spec:

\begin{lstlisting}[language=Haskell]
spec :: Spec
spec = do
  prop "notInt32Max" (forall (stream < constI32 maxBound))
  trigger "streamAdd" ...
\end{lstlisting}

Here, the function \lstinline|prop| is used to declare a property named
\lstinline|notInt32Max|, and the \lstinline|forall| combinator is used to
express that the property should hold for all elements in the stream.
The verifier must then be instructed to assume the \lstinline|notInt32Max|
property during verification so that the addition is well defined.

\subsubsection{Inductive arguments}

There are certain properties about stream programs that hold true because of
inductive reasoning that \verifier{} is nevertheless unable to prove.
For example, consider this specification that involves array indexing:

\begin{lstlisting}[language=Haskell]
sArr :: Stream (Array 2 Bool)
sArr = [array [True, False]] ++ sArr

sIdx :: Stream Word32
sIdx = [1, 0] ++ sIdx

spec :: Spec
spec = trigger "sArrIdx" (sArr .!! sIdx) [arg sArr, arg sIdx]
\end{lstlisting}

In the stream \lstinline|sArr|, each element is a length-2 array containing
\lstinline|True| and \lstinline|False|.
The \lstinline|(.!!)| combinator indexes into a stream of arrays using a stream
of indices, so \lstinline|sArr .!! sIdx| computes a stream of \lstinline|Bool|s
whose elements alternate between \lstinline|False| and \lstinline|True|
consecutively.

Each use of \lstinline|(.!!)| requires that the index value be within the
bounds of the array to be well formed, and \verifier{} discharges this
requirement as a proof goal.
Somewhat surprisingly, however, the verifier is unable to conclude that the use
of \lstinline|(.!!)| is within bounds. To see why, consider the C code that is
generated for the \lstinline|sArrIdx| trigger:

\begin{lstlisting}[language=C]
static bool sArr[1][2] = {{true, false}};
static size_t sArr_idx = 0;

static uint32_t sIdx[2] = {1, 0};
static size_t sIdx_idx = 0;

bool* sArr_get(size_t x) {
  return sArr[(sArr_idx + x) % 1];
}

uint32_t sIdx_get(size_t x) {
  return sIdx[(sIdx_idx + x) % 2];
}

bool sArrIdx_guard(void) {
  return sArr_get(0)[sIdx_get(0)];
}
\end{lstlisting}

The part of this code that poses issues for the verifier is the
\lstinline|sArrIdx_guard| function, which indexes an array using an element
from \lstinline|sIdx|.
Note that during the transition correspondence phase, the verifier creates
symbolic values for each ring buffer.
Because \lstinline|sIdx| contains symbolic values, the value that
\lstinline|sIdx_get(0)| returns is also symbolic.
Consequently, the verifier cannot determine if the returned value lies within
the bounds of \lstinline|sArr|, so it fails with an unsatisfied proof goal.

It is possible to prove by induction on the number of steps that
\lstinline|sIdx_get(0)| is within bounds by observing that the value it returns
is less than \lstinline|2|.
However, the verifier does not currently infer inductive invariants, and the
user must manually state the invariant using \lstinline|prop| so that it can be
assumed true during verification.

\subsubsection{SMT solver performance}
\label{sec:smt-solver-performance}

The performance of SMT solvers is very susceptible to small changes in how
formulas are represented.
A small change in a formula can sometimes make the SMT solver take an
arbitrarily long time to verify a proof goal, which can affect the performance
of \verifier{}.

Because \verifier{} is designed as a push-button solution, users do not have
the ability to modify the intermediate SMT formula before it is passed to the
SMT solver, or to attempt to prove it manually if the SMT solver fails to prove
it automatically.
As an alternative, users can modify the spec in a way that it generates a proof
goal that the SMT solver is better able to handle (see
\Cref{sec:floating-point} for an example), or they can use a configuration
option to alter the shape of proof goals (see \Cref {sec:configuration-options}
for details).

\section{Implementation}
\label{sec:implementation}

\verifier{} is implemented as a Haskell library that builds on top of \copilot-
and \crucible-related libraries as dependencies.
The verifier is maintained and released as part of the overall \copilot{} suite
of libraries.
\verifier{}'s source code can be installed from
Hackage~\citep{copilotVerifierHackage} or from
GitHub~\citep{copilotVerifierGitHub}.
Currently, the verifier comprises 1,854 lines of code (excluding comments and
whitespace) spread among four modules.
As part of this work, we also extended \copilotTheorem{}, increasing its total
line count by 871 lines.

The \verifier{} library has a minimal API.
In most cases, a \verifier{} user only needs to invoke the \lstinline|verify|
function.
\lstinline|verify| generates a C program from a \copilot{} specification,
compiles it to LLVM bitcode using Clang, and then verifies the behavior of the
bitcode against the specification.
After the verification process finishes, it reports whether verification
succeeded or not as output.
A typical invocation looks like this:

\begin{lstlisting}[language=Haskell]
main :: IO ()
main = do
  spec' <- reify spec
  verify mkDefaultCSettings [] "c_program_name" spec'
\end{lstlisting}

Here, \lstinline|mkDefaultCSettings| instructs Clang to compile the generated C
code using a reasonable default set of compiler flags, and \lstinline|[]|
indicates that we do not make any additional assumptions about the behavior of
the streams in the specification.
(See \Cref{sec:invariants-for-partial-operations} for an example where
verification uses an additional assumption.)

\subsection{Configuration options}
\label{sec:configuration-options}

In addition to the \lstinline|verify| function, the \verifier{} API offers a
\lstinline|verifyWithOptions| function that takes an additional set of
configuration options.
\lstinline|verifyWithOptions| is intended for users who wish to fine-tune the
verifier's behavior for particular use cases.
We describe the available configuration options below.

\subsubsection{SMT solvers}

By default, \verifier{} uses the Z3~\citep{z3} SMT solver to solve proof goals.
Users may wish to choose a different SMT solver, however, as solvers differ in
their performance as well as the types of proof goals that they can prove.
The verifier offers an \lstinline|smtSolver| setting that allows overriding the
default choice of Z3.
Aside from the Z3 solver, \verifier{} supports the use of either
CVC4~\citep{cvc4}, CVC5~\citep{cvc5}, or Yices~\citep{yices}.

\subsubsection{IEEE-754 floating-point mode}

SMT solvers have limited support for floating-point operations in general, and
there are some operations found in \copilot{} specifications that SMT solvers
do not support at all, such as transcendental functions.
As noted in \Cref{sec:floating-point}, \verifier{} works around this problem by
treating all floating-point operations as uninterpreted functions by default.
This allows \verifier{} to perform very basic equivalence checks involving
floating-point operations, but this approach is fragile in the presence of
compiler optimizations.

As an alternative, \verifier{} offers an \lstinline|smtFloatMode| setting to
allow users to interpret floating-point operations using IEEE-754 semantics
where possible.
Under this mode, the verifier will be able to reason more deeply about the
subset of floating-point operations that SMT solvers do support, such as basic
arithmetic.
This allows the verifier to be slightly more robust in the face of
floating-point--related compiler optimizations.
This approach does \emph{not} solve the problem of SMT solvers lacking support
for operations such as transcendental functions, however.
In fact, the verifier currently aborts if it encounters such an operation in
IEEE-754 mode, so this mode can only be enabled on programs that do not make
use of these operations.
We are currently evaluating alternative approaches (\Cref{sec:conclusions}).

\subsubsection{Side conditions for partial operations}

As described in \Cref{sec:partial-operations}, \verifier{} is strict about
partial operations by default: if it detects any undefined behavior
in the C code during simulation, it will cause the proof to fail.
As an alternative, \verifier{} users can enable an
\lstinline|assumePartialSideConds| setting that causes the verifier to check
that any invocation of a partial operation in the C code corresponds to an
invocation of the same partial operation in the original \copilot{} spec.
This allows users to check for ``crash-equivalence'', which may be useful for
uncovering properties about a spec beyond uses of partial operations on
undefined inputs.

\subsubsection{Logging proof goals and SMT interactions}

By default, \verifier{} tries not to overwhelm users with the information it
presents.
For this reason, when it performs a successful proof, it displays a short
summary:

\begin{lstlisting}
Translated bitcode into Crucible
Generating proof state data
Computing initial state verification conditions
Proving initial state verification conditions
Proved 5 of 5 total goals
Computing step bisimulation verification conditions
Proving step bisimulation verification conditions
Proved 14 of 14 total goals

copilot-verifier has produced a mathematical proof that
the behavior of the generated C program precisely matches
the behavior of the Copilot specification. ...
\end{lstlisting}

The output reports the number of proof goals, but not what each proof goal
means or from where in the program the goal arises.
Knowing the provenance and full details of each proof goal is vital for certain
use cases, such as when the evidence produced by \verifier{} is used as part of
an assurance case (see \Cref{sec:assurance-cases}).
For this reason, the verifier offers a \lstinline|verbosity| setting that, when
set to a high level, prints the details of each proof goal and SMT solver
interaction.
\Cref{fig:gsn-example} includes examples of what verbose proof goals look
like.

\subsection{Code style}

\verifier{} is built on top of both the \copilot{} and \crucible{} library
ecosystems.
While both libraries are written in Haskell, each one is written in very
different coding styles.
\copilot{} is largely written using simple Haskell2010 code and avoids the use
of language extensions, except where strictly necessary.
\crucible{}, on the other hand, makes heavy use of advanced Haskell features
present in the Glasgow Haskell Compiler (GHC), such as GADTs~\citep{gadts1,
gadts2}, type families~\citep{typefamilies1, typefamilies2}, data
kinds~\citep{datakinds}, and kind-indexed GADTs~\citep{kindequality}.
\verifier{}, as a tool built on top of the \crucible{} API, inherits its use of
advanced GHC features.

The \crucible{} authors describe their experience in maintaining their
particularly advanced style of Haskell code in \citet{crucible}, and our
experiences largely match theirs.
Just like in \citet{crucible}, we observe that Haskell's type system allows us
to have a high degree of confidence that certain properties of \verifier{} are
correct, such as ensuring that SMT queries that we send to the solver are
always well typed.
This, in turn, allows us to more confidently refactor the verifier's code, as
shallow bugs are usually caught by GHC's type checker.
On the other hand, we have the same difficulties that \citet{crucible} do:
namely, it can be difficult to find Haskell developers familiar with all of the
language features used in \verifier's code base, and training Haskell
developers to learn these features can be time-consuming.
We continue to evaluate the tradeoffs in the style of code used in the verifier
in an attempt to strike the right balance for our needs.

\subsection{Development process}

\verifier{} is developed as part of the overall \copilot{} framework, which
also includes an interpreter, C99 and Bluespec backends, an interactive
visualizer, a pretty-printer, and a library of stream combinators implementing
functionality generally useful to monitor systems in aerospace (e.g., temporal
logic, statistics, voting, clocks, stacks, and regular expressions).

As an intricate part of the \copilot{} project, \verifier{} is developed
following a set of rules and processes designed to meet the requirements for
Class D, Research and Technology Software, as described in NASA Software
Engineering Requirements (NPR 7150.2D)~\citep{npr71502d}.
Among other qualities, our process requires full traceability from code changes
to issues and milestones, extensive documentation of both the code and each
individual commit, following a strict process to document, confirm, and
schedule an issue before its solution is implemented, and reproducible scripts
that replicate a bug before a pull request is accepted into the mainline and
that demonstrate the bug's absence in the pull request.
This process is designed so that we can check for compliance with some of NPR
7150.2D requirements automatically~\citep{paradis2023towards}.
Additionally, \copilot{} is released on a fixed cadence (exactly every two
months) in an effort to demonstrate the project's commitment towards
reliability.

The code produced by \copilot{} has been used to support missions at higher
levels of criticality (NPR 7150.2D, Class C, Mission Support Software or
Aeronautic Vehicles, or Major Engineering/Research Facility Software).
\verifier{} is an essential piece towards making an assurance argument to
operate \copilot-generated code at that level.

\section{Related work}
\label{sec:related-work}

\copilot{} was originally based on Lustre~\citep{caspi:1987:lustre} and
LOLA~\citep{dangelo:2005:lola}.
Other similar RV frameworks include RMOR~\citep{havelund2008runtime}, which
compiles to C, and Proteus~\citep{mcclelland2021towards,mcclelland2021adding},
which compiles to C++.
\verifier{} shares similarity with other translation validation--based
compilers, such as previous efforts to translate
SIGNAL~\citep{pnueli1998translation} and
Simulink~\citep{ryabtsev2009translation} to C, the Alive2 bounded translation
validation tool for LLVM~\citep{lopes2021alive2}, and the C-to-binary
translation validation used in the seL4 operating system
kernel~\citep{2013:sewell:translation-validation}.

Most other RV frameworks do not verify their compiled code against the
specifications.
A notable exception is Lustre, which boasts a verified compiler named
\velus~\citep{Bourke2017, Bourke2019, Bourke2021}.
\velus{} is an ambitious project that verifies the Lustre compilation pipeline
within the Coq theorem prover by building on CompCert~\citep{compcert}.
This is in contrast to \verifier's translation validation approach, which only
allows verifying individually translated programs in isolation.

\velus{} first translates Lustre's stream functions into a synchronous
transition code (STC) language based on state transitions and values rather
than streams.
STC is, in turn, translated into an object-oriented language (Obc) with each
Lustre node represented as a class with its own memory, possessing a reset
method performing initialization and a step method to process the next instant
in time.
Obc is translated into CompCert's Clight, which is then translated into
assembly.
Coq is used to manually prove an equivalence at each translation step.
The proofs are aided by a memory semantics given in terms of streams of memory
trees.
The culmination of the effort is a bisimulation theorem relating the behavior
of a Lustre node and the generated assembly code.

An approach grounded in manual proof can be more complete than \verifier's
automated approach.
For instance, the difficulties that we encountered with floating-point
operations (\Cref{sec:design-tradeoffs}) could be surmounted using a Coq
formalization of floating-point
computation~\citep{boldo2011flocq,ramananandro2016unified}.
On the other hand, adapting \velus{} to build a verified Copilot compiler would
be non-trivial.
Although the core stream languages are very similar, \velus{} introduces
complexities such as the object-oriented intermediate language to accommodate
Lustre features not present in Copilot.
Alternatively, one could build a verified Copilot compiler from scratch, but in
either case, considerable resources would need to be dedicated to the effort.
Our approach to Copilot verification was influenced by the need to assure an
existing compiler for a DSL being used at NASA, and to do so under significant
resource and time constraints, hence demonstrating the applicability of the
technique to many industrial projects.

Previous versions of \copilot{}~\citep{pike2012experience,pike2013copilot}
performed limited verification of C code by compiling
with two different backends and verifying the equivalence of the generated
programs using CBMC~\citep{cbmc}.
This approach can potentially catch many of the same issues as \verifier{}, but
it does not prove that the C programs match the
original \copilot{} monitors.

Current versions of \copilot{} use property-based
testing~\citep{1997:fink:propertybased,quickcheck} for added assurance.
Although techniques such as property-based testing could catch some of the same
bugs that \verifier{} was able to detect, it would be unlikely to detect some
of the memory safety bugs uncovered by \verifier{} (some of which went
unnoticed until recently~\citep{copilotIssue:386}).

There have been efforts in \copilot{} to generate C code that includes
Hoare-logic--style contracts as function comments, where the contracts are
derived from parts of the stream program that the C function should implement.
The Frama-C static analyzer~\citep{framaC} is then used to prove that the C code
satisfies the contracts~\citep{goodloe2016challenges}.
This approach is not fully automated: manual edits are sometimes
required for the generated contracts to pass the analysis checks.

\section{Conclusions}
\label{sec:conclusions}

We have presented \verifier{}, a verifier that proves correspondences between high-level
\copilot{} monitors and low-level C code generated by the \copilot{} compiler.
Our aim was to increase confidence in the \copilot{} compiler while working
within a realistic engineering budget, which we achieved through careful design
choices and strategic use of existing tools.

Currently, \verifier{} is limited in its treatment of modeling transcendental,
floating-point functions such as sine, cosine, etc., which reduces its
applicability.
In the future, we would like to evaluate an approach where transcendental
functions are left uninterpreted, but other floating-point functionality is
modeled using IEEE-754 semantics.

In the future, we plan to use \verifier{} to construct assurance cases that are
amenable to certification.
Specifically, the NASA tool Ogma~\citep{ogmaGitHub} uses \copilot{} to produce
complete monitoring applications for NASA Core Flight System, Robot Operating
System 2, and F Prime~\citep{fprime}.
We plan to extend Ogma to leverage \verifier{} to produce evidence of assurance
for the applications generated.
In addition to demonstrating that the original \copilot{} spec matches the LLVM
bitcode, we plan to extend Ogma to facilitate checking if the behavior of the C
program is altered by the choice of compiler or compiler flags, via a
combination of lightweight formal methods (fuzzing, differential testing,
etc.).

\section*{Acknowledgements}
\label{sec:acknowledgements}

We thank the reviewers for their feedback and suggestions.
This manuscript has been co-authored by Ivan Perez, an employee of KBR under
Prime Contract No. 80ARC020D0010 with the NASA Ames Research Center.
Any opinions, findings, and conclusions or recommendations expressed in this
material are those of the authors and do not necessarily reflect the views,
either expressed or implied, of any of the funding organizations.
The United States Government retains, and by accepting the article for
publication, the publisher acknowledges that the United States Government
retains, a non-exclusive, paid-up, irrevocable, worldwide license to publish or
reproduce the published form of this work, or allow others to do so, for United
States Government purposes.

\clearpage
\bibliographystyle{ACM-Reference-Format}
\bibliography{bibliography,bibliography-deanonymized}


\end{document}